\documentclass[fleqn,usenatbib]{mnras}

\usepackage{newtxtext,newtxmath}
\usepackage[T1]{fontenc}
\usepackage{ae,aecompl}
\usepackage{booktabs}

\usepackage{rotfloat}
\usepackage[tableposition=top]{caption}
\usepackage{pdflscape}

\usepackage{graphicx}	
\usepackage{amsmath}	
\usepackage{amssymb}	
\usepackage{comment}

\newcommand{\dd}{{\rm d}}
\newcommand{\vk}{v_{\rm k}}

\newcommand{\vsys}{v_{\rm sys}}
\newcommand{\MNS}{M_{\rm NS}}
\newcommand{\MHe}{M_{\rm He}}
\newcommand{\tmerge}{t_{\rm merge}}
\newcommand{\taumerge}{\tau_{\rm merge}}
\newcommand{\dtravel}{d_{\rm travel}}
\newcommand{\degree}{\ifmmode {^{\circ}\ }\else$^{\circ}$\fi}

\newcommand{\Msun}{\ifmmode {M_{\odot}}\else${M_{\odot}}$\fi}
\newcommand{\Rsun}{\ifmmode {R_{\odot}}\else${R_{\odot}}$\fi}
\newcommand{\Porb}{\ifmmode {P_{\rm orb}}\else${P_{\rm orb}}$\fi}

\title[DNS Merger Times, Velocities and Travel Distances]{Double Neutron Star Formation: Merger Times, Systemic Velocities, and Travel Distances}

\author[Andrews \& Zezas]{
Jeff J.~Andrews,$^{1,2,3}$\thanks{E-mail: andrews@physics.uoc.gr} Andreas Zezas$^{1,2,4}$
\\
$^1$ Foundation for Research and Technology -- Hellas, IESL, Voutes, 71110 Heraklion, Greece \\
$^2$ Physics Department \& Institute of Theoretical \& Computational Physics, University of Crete, 71003 Heraklion, Crete, Greece \\
$^3$ Niels Bohr Institute, Blegdamsvej 17, 2100 K\o benhavn \O, Denmark \\
$^4$ Center for Astrophysics, Cambridge, MA, USA
}

\date{Accepted XXX. Received YYY; in original form ZZZ}

\pubyear{2019}

\begin{document}
\label{firstpage}
\pagerange{\pageref{firstpage}--\pageref{lastpage}}
\maketitle

\begin{abstract}
The formation and evolution of double neutron stars (DNS) have traditionally been studied using binary population synthesis. In this work, we take an alternative approach by focusing only on the second supernova (SN) forming the DNS and the subsequent orbital decay and merger due to gravitational wave radiation. Using analytic and numerical methods, we explore how different NS natal kick velocity distributions, pre-SN orbital separations, and progenitor He-star masses affect the post-SN orbital periods, eccentricities, merger times, systemic velocities, and distances traveled by the system before merging. Comparison with the set of 17 known DNSs in the Milky Way shows that DNSs have pre-SN orbital separations ranging between 1 and 44 \Rsun. Those DNSs with pre-SN separations $\sim$1 \Rsun\ have merger time distributions that peak $\sim$10-100 Myr after formation, regardless of the kick velocity received by the NS. These DNSs are typically formed with systemic velocities $\sim$10$^2$ km s$^{-1}$ and may travel $\sim$1-10 kpc before merging. Depending on progenitor mass of the second-born NS, the short merger time can account for the $r$-process enrichment observed in compact stellar systems such as ultra-faint dwarf galaxies. For Milky Way-mass galaxies only DNSs with the tightest pre-SN orbits and large kick velocities ($\gtrsim$10$^2$ km s$^{-1}$) can escape. However, those DNSs that do escape may travel as far as $\sim$Mpc before merging, which as previous studies have pointed out has implications for identifying the host galaxies to short gamma ray bursts and gravitational wave events.
\end{abstract}


\section{Introduction}
\label{sec:intro}

The recent discovery by LIGO/Virgo of the merger of a double neutron star system (DNS), GW170817, has opened a window into previously untested physics \citep{ligo_detection}. Among others, these include the NS equation of state \citep{radice18}, the origin of $r$-process elements in the Universe \citep{pian17, drout17}, and a new, independent measure of the Hubble constant \citep{ligo_hubble}. Due to their importance for a broad range of physics, much effort has been spent uncovering DNS formation and evolution since the detection of the first DNS, the Hulse-Taylor system \citep{hulse75, flannery75, de_loore75a, de_loore75b, srinivasan82, bhattacharya91, tauris06}. 

There is now a consensus on the broad evolutionary scenario forming DNS \citep[see][and references therein]{tauris17}. DNS begin their evolution as two high mass stars on the main sequence. The more massive star evolves first into a NS, which may receive a kick at birth as high as several hundred km s$^{-1}$. After some time, the secondary will also evolve off the main sequence, expanding until it overfills its Roche lobe, thereby losing its hydrogen envelope. The binary, now comprised of a NS with a ``naked'' helium star companion, may undergo a further phase of mass transfer, the so-called Case BB mass transfer phase \citep{delgado81,dewi02,ivanova03,tauris13}, in which the helium star also overfills its Roche lobe. Eventually, the secondary will undergo core collapse, and if its own natal kick does not unbind the system, form a DNS. For systems with tight enough orbits, the angular momentum and energy dissipation due to gravitational wave radiation will cause the system to merge within a Hubble time.

Analysis has shown that modest variations in the initial binary distributions do not propagate into large deviations for the compact object mergers \citep{deMink15}; however, the uncertainties in the details of the binary evolution pathway outlined above are numerous, and the approximations taken can be extreme. As one example, it is currently unknown to what extent the NSs in DNSs are formed through the canonical collapse of a degenerate Fe-core, electron captures onto a semi-degenerate ONe core \citep[electron capture SN;][]{nomoto84, jones16, gessner18}, or the collapse of a stripped-core in which both H and He have been lost due to prior mass transfer \citep{tauris13, tauris15}. This uncertainty, which applies to the formation of the first NS as well as the second \citep{schwab10,giacobbo19}, affects the magnitude of the NS natal kicks and hence the subsequent evolution of the system.

Any viable model predicting the properties of merging DNSs detectable by gravitational wave observatories must also account for the growing population of Galactic DNSs. Historically, the approach to studying the properties of Galactic DNS, including their merger times and systemic velocities, has been to use binary population synthesis \citep[BPS;][]{tutukov93, portegies_zwart98, belczynski99, voss03, belczynski08, oslowski11, andrews15, chruslinska18, belczynski18, vigna-gomez18}. This is a Monte Carlo method in which one makes random draws of initial binary parameters from observationally motivated distributions of binary star populations, then evolves this synthetic population forward using our best understanding of the physics involved to obtain a population of DNSs. As the population of observed Galactic DNS has grown, so has our understanding of the details of DNS formation. Currently there are 17 known or suspected field DNSs in the Galaxy (we ignore DNSs in globular clusters; for a complete list of field DNSs, see \citet{tauris17} and references therein, as well as the newly detected systems by \citet{lynch18}, \citet{cameron18}, and \citet{stovall18}) and one extragalactic DNS merger event detected. With instrument upgrades to the LIGO/Virgo network \citep{martynov16} and new radio observatories coming online \citep[e.g., SKA;][]{tauris15b, levin17}, these numbers will likely multiply in the coming years. 

The distribution of DNS orbital periods, eccentricities, merger times, systemic velocities, and distances traveled before merger have all been previously analyzed using BPS \citep[e.g.,][]{portegies_zwart98,belczynski02a, belczynski02b}. In this work, we use an alternative approach to BPS to study DNS mergers. We ignore the uncertain binary evolution prior to the second SN forming the DNS and instead focus only on the dynamical effects of the second SN and the subsequent orbital decay due to gravitational wave radiation. Because the calculations accounting for these dynamical effects are exact, our results depend only on the joint assumptions that the SN forming the second NS occurs effectively instantaneously in a circular pre-SN orbit and that the natal kick it receives is isotropically directed with either a single, prescribed magnitude or a magnitude drawn from a Maxwellian distribution. Using analytic and semi-analytic approaches, we derive the delay time, systemic velocity, and travel distance distributions of DNSs for a range of SN kick distributions and pre-SN orbital parameters. This approach allows us to focus on regions of the parameter space that are poorly populated with Monte Carlo random sampling, but may be important to the overall population in the Universe.

Our work builds off of the previous studies by \citet{kalogera96} who uses a similar approach to follow the orbital evolution of a high mass binary through the first SN in the system and subsequent tidal circularization using a Maxwellian kick distribution. Later work by \citet{kalogera00b} and \citet{kalogera01} focuses on the evolution of DNSs through the second kick, but this only analyzed the case of isotropic kicks with a single magnitude. Other notable work in this direction include \citet{kalogera00} who studies the spin-orbit misalignment angle of NS-black hole binaries. In this work, we comprehensively analyze the effects of both isotropic, single-magnitude kicks and Maxwellian kicks and compare with the current, expanded sample of Galactic DNS.

In Section \ref{sec:dynamics}, we describe our formalism for calculating the post-SN characteristics and the orbital decay from gravitational wave radiation. These equations greatly simplify under the assumption of no kick received by the NS at birth, which we describe in Section \ref{sec:blaauw}. We follow this with a description of the orbital characteristics for a limited grid in Section \ref{sec:orbital_characteristics}. Applying gravitational wave radiation to these orbital distributions gives us distributions of merger times which we discuss in Section \ref{sec:merger_time} and systemic velocities and distances traveled before merging which we discuss in Section \ref{sec:distance_velocity}. We discuss the qualitative comparison with various observational data sets in Section \ref{sec:discussion}, and we conclude in Section \ref{sec:conclusions}.

\section{Supernova and GR Orbital Decay Dynamics}
\label{sec:dynamics}

In this section we derive the distributions of post-SN orbits following the method of \citet{kalogera96} for both isotropic, single-magnitude kicks and isotropic kicks following a Maxwellian distribution. We then use the equations of \citet{peters64} to determine the orbital evolution and merger time of systems due to gravitational wave radiation. Where possible, we ensure the accuracy of the results throughout this paper by testing against Monte Carlo simulations of SN kicks.

\subsection{Supernova Orbital Dynamics}
\label{sec:SN_orbit}

Observationally, it is well established that NSs receive a kick at birth \citep{fryer97,hobbs05}. For kick velocities of hundreds of km s$^{-1}$, many binary systems may be disrupted, depending on the parameters of the system \citep{tauris98}. However, for the right combination of kick direction and magnitude, many systems may remain bound, but with post-SN orbits significantly altered. Furthermore, the surviving DNS {\it system} receives a kick, $V_{\rm sys}$, which may be similar in magnitude to the NS kick \citep{wex00}.

Although NS natal kicks have been historically modeled with an isotropic, Maxwellian distribution with a dispersion velocity, $\sigma_k$, recent work has shown that this distribution may not best represent the observations \citep{bray16, janka17, bray18}. To create broadly applicable results, we explore both isotropic kick distributions of a single magnitude as well as isotropic, Maxwellian kick distributions. 

We adopt the kick magnitude prescription by \citet{bray18} to represent the isotropic, single-magnitude kicks that NS receive at birth:
\begin{equation}
V_{\rm k} = \left| 100 \frac{M_{\rm He} - M_{\rm NS}}{M_{\rm NS}} - 170 \right|\ {\rm km}\ {\rm s}^{-1}, \label{eq:V_kick_BE}
\end{equation}
where $M_{\rm He}$ and $M_{\rm NS}$ are the pre-SN helium star and NS masses, respectively. Note that we refer to the NS progenitor as a helium star, even though multiple phases of mass transfer may have removed most of the helium envelope prior to core collapse \citep{tauris13}. The absolute value in Equation \ref{eq:V_kick_BE} is necessary since without it, the kick magnitude would be negative for sufficiently low helium star masses \citep[see discussion in][]{bray18}. For reference in Table~\ref{tab:V_kick} we provide the kick velocities from this prescription for the He-star masses used throughout this work.

\begin{table}
  \centering
  \caption{The kick velocity applied to newly born 1.4 \Msun\ NSs with different He-star progenitor masses according to the prescription from \citet{bray18} provided in Equation \ref{eq:V_kick_BE}.}
  \label{tab:V_kick}
  \begin{tabular}{lcccc}
  \hline
  $M_{\rm He}$ ($M_{\odot}$) & 1.5 & 2 & 3 & 5 \\
  \hline
  $V_{\rm k}$ (km s$^{-1}$) & 163 & 127 & 56 & 87\\
  \hline
  \end{tabular}
\end{table}

\begin{figure}
	\includegraphics[width=0.9\columnwidth]{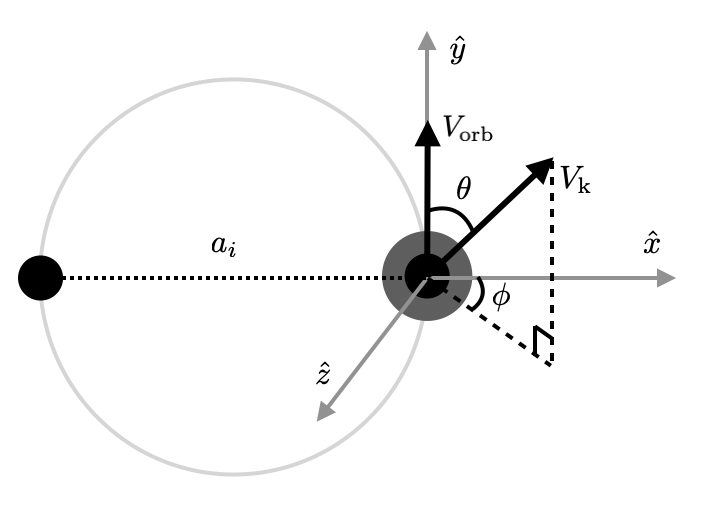}
    \caption{ The pre-SN orbital plane lies in the $\hat{x}$-$\hat{y}$ plane (the plane of the page), in the inertial frame of the companion, but with the coordinate system centered on the collapsing star. The He-star progenitor of the NS is initially moving with the orbital velocity $V_{\rm orb}$ in the positive $\hat{y}$ direction. $\theta$ is a polar angle, defined with respect to the orbital velocity of the collapsing star, while $\phi$ is an azimuthal angle, measured in the $\hat{x}$-$\hat{z}$ plane. These two angles follow \citet{kalogera96} \citep[rather than][]{hills83} as this choice removes any dependence of $\phi$ on the post-SN orbital separation. }
    \label{fig:SN_diagram}
\end{figure}

Following \citet{kalogera96}, we express the post-SN orbit in non-dimensional variables $e$ and $\alpha$, where $e$ is the orbital eccentricity and $\alpha$ is the non-dimensional post-SN orbital separation, $a_f$, expressed in terms of the pre-SN orbital separation, $a_i$:
\begin{equation}
\alpha = \frac{a_f}{a_i}. \label{eq:alpha_def}
\end{equation}
Again following \citet{kalogera96}, the post-SN systemic velocity, NS kick velocity, and SN kick Maxwellian dispersion velocity may similarly be expressed in non-dimensional forms:
\begin{eqnarray}
\vsys &=& \frac{V_{\rm sys}}{V_{\rm orb}} \\
\vk &=& \frac{V_{\rm k}}{V_{\rm orb}} \\
\xi &=& \frac{\sigma_k}{V_{\rm orb}},
\end{eqnarray}
where $V_{\rm orb}$ is the pre-SN orbital velocity:
\begin{equation}
V_{\rm orb} = \sqrt{ \frac{\mathcal{G} (\MHe + M_2)}{a_i} }. 
\end{equation}
In this equation, $M_2$ is the companion mass. \citet{kalogera96} also provide the post-SN orbital parameters for a kick with a magnitude directed along a polar angle $\theta$ measured with respect to the orbital motion vector and an azimuthal angle $\phi$ that is perpendicular to the orbital axis. Figure \ref{fig:SN_diagram} shows the orientation of $\theta$ and $\phi$ with respect to the pre-SN orbital plane. The post-SN separation $\alpha$ and eccentricity $e$ are
\begin{eqnarray}
\alpha &=& \frac{\beta}{2\beta - \vk^2 - 1 - 2 \vk\cos \theta} \label{eq:alpha} \\
1 - e^2 &=& \frac{1}{\alpha \beta} \left( 2 \beta - \frac{\beta}{\alpha} - \vk^2 \sin^2 \theta \cos^2 \phi \right),  \label{eq:ecc}
\end{eqnarray}
where $\beta$ is the ratio of the post-SN to pre-SN total system masses:
\begin{equation}
\beta = \frac{\MNS + M_2}{\MHe + M_2}.
\end{equation}
Since we focus on the second SN in DNS formation, the companion has already evolved into a NS: $M_2$$=$$\MNS$$=$1.4 $\Msun$. The mass of the He-star progenitor to the NS, $\MHe$, is a variable we explore throughout this work, but it is limited to be greater than $\MNS$, so that $\beta < 1$. While some NSs in DNSs have masses somewhat lower than 1.4 \Msun, we expect any variation to only provide a perturbative effect on our results.

\subsubsection{Isotropic, Single-Speed NS Kicks}

We first look at kick distributions with a single magnitude kick directed isotropically on the sky, so:
\begin{eqnarray}
P(\cos \theta) &=& \frac{1}{2} \label{eq:P_cos_theta} \\
P(\phi) &=& \frac{1}{2 \pi}. \label{eq:P_phi}
\end{eqnarray}
Following \citet{kalogera00b} and \citet{kalogera01} we perform a Jacobian transformation to convert the distribution of angles $\theta$ and $\phi$ into the post-SN orbital properties, $\alpha$ and $e$:
\begin{eqnarray}
P(\alpha, e) &=& P(\cos \theta, \phi) 
	\left| 
		\begin{array}{cc}
        \frac{\partial \cos \theta}{\partial \alpha} & \frac{\partial \cos \theta}{\partial e} \\
        \frac{\partial \phi}{\partial \alpha} & \frac{\partial \phi}{\partial e}
		\end{array}
	\right| \nonumber \\
    &=& P(\cos \theta) P(\phi) \left| \frac{\partial \cos \theta}{\partial \alpha} \right| \left|\frac{\partial \phi}{\partial e} \right|. \label{eq:P_alpha_ecc_jacobian}
\end{eqnarray}
The simplification is made because $\cos \theta$ and $\phi$ are independent parameters and Equation \ref{eq:alpha} shows that $\alpha$ has no dependence on $\phi$. After calculating the partial derivatives in Equation \ref{eq:P_alpha_ecc_jacobian} and inserting the probabilities in Equations \ref{eq:P_cos_theta} and \ref{eq:P_phi}, we arrive at a closed form expression for $P(\alpha, e)$:
\begin{equation}
P(\alpha, e) = \frac{e}{\pi \alpha \vk} \sqrt{\frac{\beta^3}{C_1 C_2}}, \label{eq:P_alpha_ecc_single}
\end{equation}
where
\begin{eqnarray}
C_1 &=& 2 - \frac{1}{\alpha} - \alpha(1-e^2) \label{eq:C_1} \\
C_2 &=& 4\vk^2 - \left(2\beta - \frac{\beta}{\alpha} - \vk^2 - 1\right)^2 - 4\beta C_1.
\end{eqnarray}

Equation \ref{eq:P_alpha_ecc_single} has two singularities. The first occurs when $C_1 = 0$, which translates to $\alpha=(1\pm e)^{-1}$. When adding the restriction that $P(\alpha, e)$ must be real, this singularity produces the constraint that $(1+e)^{-1} < \alpha < (1-e)^{-1}$. This constraint, which was first shown by \citet{flannery75} based on energy and angular momentum conservation arguments \citep[and further described by][]{kalogera96} can be understood intuitively: the binary's separation at the moment of SN must be greater than the periastron separation, $\alpha(1-e)$, and less than the apastron separation, $\alpha(1+e)$, of the post-SN orbit. It is worth noting further that this constraint is true so long as the assumptions of a circular pre-SN orbit and an instantaneous kick are valid; this constraint holds even for non-isotropic SN kicks \citep[see also][who derive a form of this constraint for SN in an elliptical pre-SN orbit]{fryer97}.

The second singularity of Equation \ref{eq:P_alpha_ecc_single} occurs when $C_2=0$, which translates to:
\begin{equation}
e^2 = 1 - \frac{\left[ \alpha (v_k^2 - 2 \beta - 1) + \beta \right]^2}{4 \alpha^3 \beta}. \label{eq:singularity_2}
\end{equation}
This singularity is an angular momentum constraint and corresponds to the maximum post-SN orbital eccentricity allowed for a given $\alpha$ when $\phi$ is equal to an integer multiple of $\pi$.

\begin{figure}
	\includegraphics[width=0.99\columnwidth]{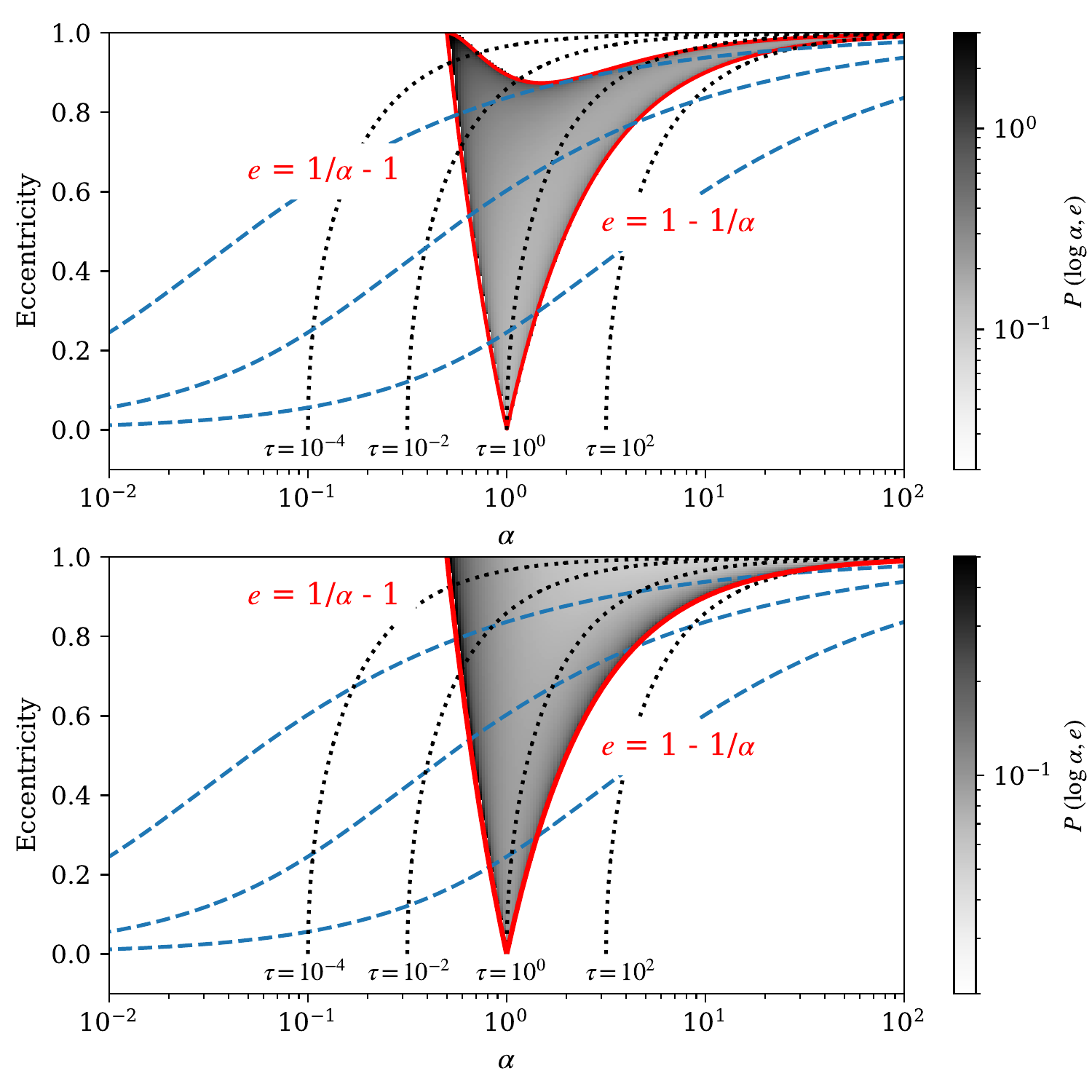}
    \caption{ The bivariate distribution, $P(\alpha, e)$. The top panel shows the resulting distribution for isotropic, single-magnitude kicks for $\beta=0.6$ and $v_{\rm k}=1$, while the bottom panel shows the corresponding distribution for isotropic Maxwellian kicks with $\beta=0.6$ and $\xi=1$. Red lines show the analytic limits on these distributions, determined by the singularities described in Section \ref{sec:SN_orbit}. Black dotted lines indicate regions of constant (non-dimensional) merger time in this two-dimensional space (see Equation \ref{eq:tau_definition} for a definition), whereas blue dashed lines show the path that systems follow as they evolve from right to left in this space due to gravitational wave radiation as the orbits circularize and decay. }
    \label{fig:alpha_ecc}
\end{figure}

In the top panel of Figure \ref{fig:alpha_ecc}, we provide the bivariate distribution of post-SN orbital parameters from Equation \ref{eq:P_alpha_ecc_single}, as well as the singularities bounding this distribution highlighted in red. Note that, for our choice of $\beta$ and $v_k$ all values of eccentricity between zero and unity are possible. For larger or smaller $v_k$, the singularity in Equation \ref{eq:singularity_2} may produce more restrictive constraints on possible post-SN orbital eccentricities.

We also wish to calculate the post-SN systemic velocities, which are provided by \citet{kalogera96} \citep[see also][]{brandt95}:
\begin{equation}
v_{\rm sys}^2 = \kappa_1 + \kappa_2 \frac{2\alpha - 1}{\alpha} - \kappa_3 \frac{v_{\rm k} \cos \theta + 1}{\sqrt{\beta}},
\end{equation}
where $\kappa_1$, $\kappa_2$, and $\kappa_3$ are unitless functions of the pre- and post-SN masses of the components:
\begin{eqnarray}
\kappa_1 &=& \frac{\MHe^2}{(\MHe + M_2)^2} \label{eq:kappa_1} \\
\kappa_2 &=& \frac{\MNS^2}{(\MHe + M_2) (\MNS + M_2)} \label{eq:kappa_2} \\
\kappa_3 &=& 2 \sqrt{\kappa_1 \kappa_2}. \label{eq:kappa_3}
\end{eqnarray}
Calculating distributions of $v_{\rm sys}$, which we explicitly show in Section \ref{sec:distance_velocity}, requires making the necessary Jacobian transformation from $P(\alpha, e)$.

\begin{figure*}
	\includegraphics[width=\textwidth]{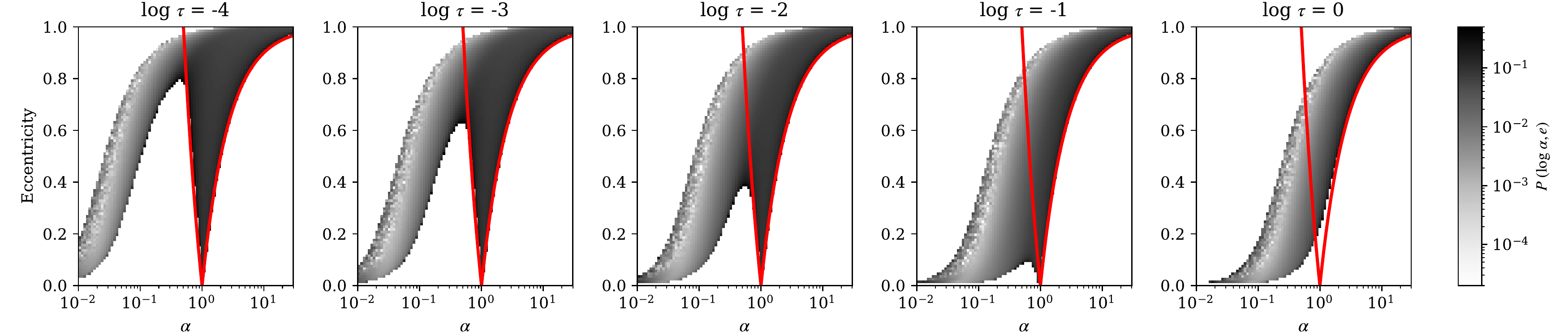}
    \caption{ The distribution of $\alpha$ and $e$ after the some time, $\tau$, has passed allowing gravitational wave radiation to circularize and shrink the orbit. As time progresses, gravitational wave radiation continuously peels off layers of the $P(\alpha, e)$ distribution, causing DNSs to merge. As in Figure \ref{fig:alpha_ecc}, we use $\beta$$=$0.6 and $\xi$$=$1 to generate this distribution. }
    \label{fig:alpha_ecc_GW}
\end{figure*}

\subsubsection{Isotropic, Maxwellian NS Kicks}

An isotropic, Maxwellian distribution of SN kicks can be thought of as a superposition of isotropic, single-magnitude kicks. This distribution has been described by \citet{kalogera96} who provides the corresponding post-SN bivariate distribution:
\begin{eqnarray}
P(\alpha, e) &=& \left( \frac{\beta}{2 \pi \xi^2} \right)^{3/2} 
\frac{2 \pi e}{\sqrt{\alpha (1-e^2)}}
\left[ \left( \alpha - \frac{1}{1+e} \right) \left( \frac{1}{1-e} - \alpha \right) \right]^{-1/2} \nonumber \\
 & & \times\ \exp \left[ - \frac{1}{2\xi^2} \left( \beta \frac{2\alpha - 1}{\alpha} + 1 \right) \right]
I_0 \left[ \frac{\sqrt{\beta \alpha (1-e^2)}}{\xi^2} \right], \label{eq:P_ecc_alpha}
\end{eqnarray}
where $I_0$ is the zeroth order modified Bessel function of the first kind. We refer the reader to \citet{kalogera96} for a derivation of this distribution. Note that Equation \ref{eq:P_ecc_alpha} can be numerically challenging to evaluate for $\xi$$\lesssim$0.1, since in the limit of small $\xi$, the exponential term trends toward zero while the modified Bessel function term approaches infinity. For several of the integrals throughout this work, double precision calculation of Equation \ref{eq:P_ecc_alpha} was insufficient.

Examination of Equation \ref{eq:P_ecc_alpha} shows that a singularity in the bivariate distribution in $\alpha$ and $e$ for Maxwellian kicks exists at $\alpha = (1\pm e)^{-1}$, identical to the corresponding single-magnitude kicks. However, unlike for single-magnitude kicks, since a Maxwellian distribution has non-zero probability for all $v_{\rm k}>0$, no second singularity exists.

Greyscale contours in the bottom panel of Figure \ref{fig:alpha_ecc} show the bivariate orbital separation and eccentricity distribution, for fiducial values of $\beta=0.6$ and $\xi=1$. As can be determined from these limits and seen in both panels of Figure \ref{fig:alpha_ecc}, the post-SN orbital separation can be no less than half the initial SN orbital separation but may extend to arbitrarily large separations, so long as the eccentricity is sufficiently close to unity that $e>1-\alpha^{-1}$.

Throughout this work, for Maxwellian kicks we will extensively use Equation \ref{eq:P_ecc_alpha} for $P(\alpha, e)$ above as well as the trivariate distribution, $P(\alpha, e, \vsys)$ also provided by \citet{kalogera96}:
\begin{eqnarray}
P(\alpha, e, \vsys) &=& \left( \frac{\beta}{2 \pi \xi^2} \right)^{3/2} 
\frac{4 e \vsys}{\kappa_3 \alpha (1-e^2)} \nonumber \\
& & \times\ \left[ \left( \alpha - \frac{1}{1+e} \right) \left( \frac{1}{1-e} - \alpha \right) \right]^{-1/2} \nonumber \\
& & \times\ \exp \left[ - \frac{1}{2 \xi^2} \left( \beta \frac{2 \alpha -1}{\alpha} + 1 \right) \right] \nonumber \\
& & \times\ \exp \left[ \frac{\sqrt{\beta}}{\kappa_3 \xi^2} \left( \kappa_1 + \kappa_2 \frac{2 \alpha - 1}{\alpha} - \vsys^2 \right) \right] \nonumber \\
& & \times\ \left\{ 1 - \frac{\left[ \kappa_1 + \kappa_2 (2\alpha -1)/\alpha - \vsys^2 \right]^2}{\kappa_3^2 \alpha (1 - e^2)} \right\}^{-1/2}, \label{eq:P_alpha_e_vsys}.
\end{eqnarray}

\subsection{Orbital Decay due to Gravitational Wave Radiation}

\begin{figure*}
	\includegraphics[width=1.0\textwidth]{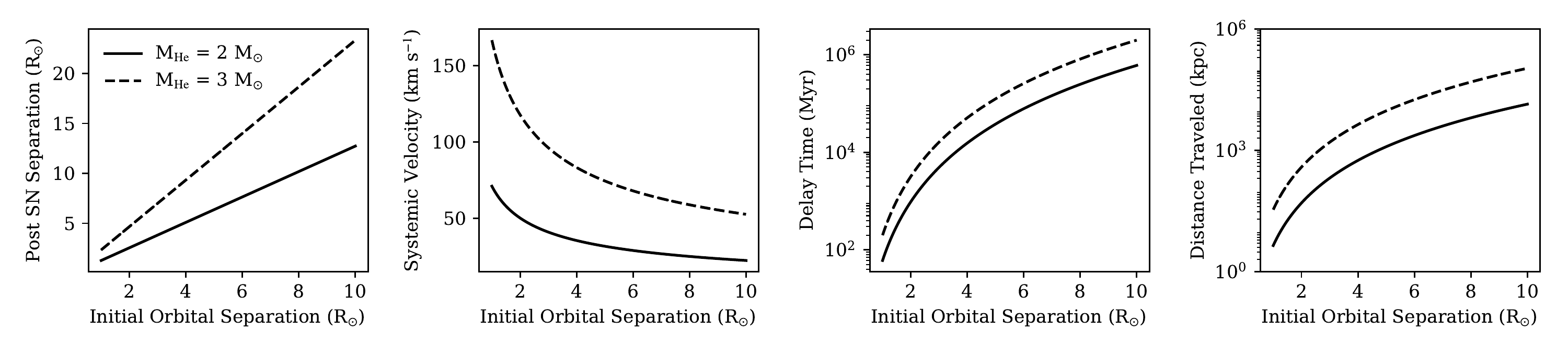}
    \caption{ With the Blaauw kick (no kick is applied to the newborn NS), the orbit evolves only due to instantaneous mass loss. As a function of the initial orbital separation, we show the post-SN orbital separation (far left panel), systemic velocity (middle left panel), delay time before merging due to gravitational wave radiation (middle right panel), and distance traveled before merging (far right panel). Due to the orbital expansion upon mass loss, and the strong dependence of the orbital separation on the merger time, only the tightest pre-SN orbits merge within a Hubble time in this limit. In an isolated environment, such systems may travel a few kpc to as much as a Mpc before merging. For a Milky Way-like galaxy, with typical escape velocities of hundreds of km s$^{-1}$ \citep{piffl14}, these velocities are too small for the DNS to escape the galaxy.}
    \label{fig:blaauw}
\end{figure*}

Those systems that survive the second core collapse to form a DNS are quasi-stable systems; both stars in the binary are completely evolved, and in typical circumstances the only effect one needs to account for is the loss of orbital energy and angular momentum by gravitational wave radiation which may ultimately lead to the system's merger. In non-dimensional units, the classical expression from \citet{peters64} for the merger time ($\taumerge$) of two point-like masses (a good approximation for DNSs, since the NS radius is many orders of magnitude smaller than the orbital separation) due to gravitational wave radiation-induced orbital decay scales with the orbital separation to the fourth power:
\begin{equation}
\taumerge = \frac{48}{19} H(e) \alpha^4, \label{eq:tau_merge}
\end{equation}
where $H(e)$ is a non-linear function of eccentricity that must be solved numerically:
\begin{eqnarray}
H(e) &=& \left(1-e^2\right)^4 e^{-48/19} \left( 1 + \frac{121}{304} e^2 \right)^{-3480/2299} \nonumber \\
 & & \times \int_0^e \dd e' \frac{e'^{29/19} \left[1 + (121/304) e'^2\right]^{1181/2299}}{(1-e'^2)^{3/2}}. \label{eq:H_e}
\end{eqnarray}
A closed-form approximation for $H(e)$ exists which is accurate to $\approx$2\% or better for $e<$0.75:
\begin{equation}
H(e) \approx \frac{19}{48} (1-e^2)^{7/2} \label{eq:H_e_low}.
\end{equation}
Combining Equations \ref{eq:tau_merge} and \ref{eq:H_e_low} shows that $\taumerge$ is defined so that a circular binary with $\alpha=1$ has a merger time equal to unity. To transform from non-dimensional to physical units we use the conversion:
\begin{equation}
\tau = \frac{4 \beta_G}{a_i^4} t, \label{eq:tau_definition}
\end{equation}
where $\beta_G$ is a non-linear function of the binary component masses provided by \citet{peters64}:
\begin{equation}
\beta_G = \frac{64}{5} \frac{\mathcal{G}^3 \MNS M_2 (\MNS + M_2)}{c^5}.
\end{equation}

In Figure \ref{fig:alpha_ecc}, we show lines (black, dotted) of constant, non-dimensional merger time, computed using Equation \ref{eq:tau_merge}. This figure shows that by a few $\tau$ the only systems that remain unmerged are those formed in the high eccentricity tail of the post-SN orbital distribution, near the $\alpha = (1-e)^{-1}$ limit.

In addition to the merger time, we are also interested in how the orbits evolve due to gravitational wave radiation for those DNSs that have not yet merged. The coupled, first order, ordinary differential equations dictating the secular evolution of orbits from \citet{peters64} can also be expressed in non-dimensional units:
\begin{eqnarray}
\left< \frac{\partial \alpha}{\partial \tau} \right> &=& - \frac{1}{4} \frac{1}{\alpha^3} \frac{1}{(1-e^2)^{7/2}} \left( 1 + \frac{73}{24}e^2 + \frac{37}{96}e^4 \right) \label{eq:dalpha_dtau} \\
\left< \frac{\partial e}{\partial \tau} \right> &=& - \frac{19}{48} \frac{1}{\alpha^4} \frac{e}{(1-e^2)^{5/2}} \left( 1 + \frac{121}{304} e^2 \right) \label{eq:decc_dtau}
\end{eqnarray}

From Equations \ref{eq:dalpha_dtau} and \ref{eq:decc_dtau} it can be seen that the evolution of eccentricity and orbital separation are strongly correlated. In Figure \ref{fig:alpha_ecc}, we show as blue, dashed lines the path traveled by systems in $\alpha-e$ space as the orbits simultaneously circularize and decay due to gravitational wave radiation. 

Obviously, decay due to gravitational wave radiation implies that the characteristics of a population of DNSs observed today will differ from the characteristics of that population at birth; exactly how depends on the age of that population. In the equations below, we use the subscripted $\alpha_{\rm o}$ and $e_{\rm o}$ to denote the orbital separation and eccentricity observed today. More generally, one can marginalize over the age distribution of DNSs, $P(\tau)$ (when $\tau$ is much greater than the formation time of the DNS, typically a few tens of Myr, $P(\tau)$ can be reasonably approximated as the star formation history of the population being studied):
\begin{equation}
P(\alpha_{\rm o}, e_{\rm o}) = \int \dd \tau'\ P(\alpha, e)\ J_{\rm GW} P(\tau') \label{eq:P_alpha_ecc_GR}
\end{equation}
where $J_{\rm GW}$ is the determinant of the Jacobian transformation matrix between the orbital parameters today, and the parameters ($\alpha, e$) some time, $\tau$, in the past, just after the second SN:
\begin{equation}
J_{\rm GW} = 
	\left| 
		\begin{array}{cc}
        \frac{\partial \alpha}{\partial \alpha_{\rm o}} & \frac{\partial \alpha}{\partial e_{\rm o}} \\
        \frac{\partial e}{\partial \alpha_{\rm o}} & \frac{\partial e}{\partial e_{\rm o}}
		\end{array}
	\right|_{\tau=\tau'}.
\end{equation}
Since $a_{\rm o}$ and $e_{\rm o}$ must be calculated by integrating backwards in time Equations \ref{eq:dalpha_dtau} and \ref{eq:decc_dtau}, which do not have closed-form solutions, the components of $J_{\rm GW}$ must each be evaluated numerically.

Figure \ref{fig:alpha_ecc_GW} shows how the bivariate distribution in $\alpha-e$ space evolves over time, calculated from Equation \ref{eq:P_alpha_ecc_GR}. Even at relatively small $\tau$, the effects of gravitational wave radiation on the distribution are apparent; the highest eccentricity and shortest orbital period systems circularize and decay. As time passes, gravitational wave radiation peels off systems from the $P(\alpha, e)$ distribution at progressively larger orbital separations, causing DNSs to merge. By $\tau$$=$1, the only systems remaining are those close to the $(1-e)^{-1}$ limit on $\alpha$.

\section{Blaauw Kick}
\label{sec:blaauw}

When the NS receives no kick at birth, the orbit is only affected by the (assumed to be instantaneous and symmetric) mass loss of the collapsing star. This paradigm is denoted the Blaauw kick after \citet{blaauw61} who first studied the dynamics of a binary orbit under this model as an explanation for run-away O- and B-type stars \citep[see also][]{boersma61}. Although the evidence that DNSs receive natal kicks is overwhelming \citep[e.g.,][]{fryer97}, it is still worth investigating the dynamics of the Blaauw kick since it provides a reasonable estimate for the evolution of systems in which the kick velocity is much smaller than the orbital velocity ($v_{\rm k}$$<<$$1$). Indeed, observations suggest that at least some DNSs are likely to be formed with small kicks \citep[e.g.,][]{van_den_heuvel07}. In this limit, the orbital mechanics simplify substantially.

In terms of the $\beta$ and the $\kappa$ expressions defined in Equations \ref{eq:kappa_1}-\ref{eq:kappa_3}, the relevant non-dimensional post-orbit variables become:
\begin{eqnarray}
\alpha &=& \frac{\beta}{2 \beta - 1} \\
e &=& \frac{1-\beta}{\beta} \label{eq:blaauw_e} \\
\vsys^2 &=& \kappa_1 + \frac{\kappa_2}{\beta} - \frac{\kappa_3}{\sqrt{\beta}}. \label{eq:blaauw_vsys}
\end{eqnarray}
For $\beta<0.5$, $\alpha$ becomes negative and $e$ must be larger than unity; physically, this means that if a binary instantaneously loses more than half its total mass, it becomes unbound, a result first obtained by \citet{blaauw61}. Assuming 1.4 \Msun\ NSs, $\MHe$ must be less than 4.2 \Msun\ for the binary to survive without a kick.

In the left two panels of Figure \ref{fig:blaauw}, we show the post-SN orbital separation and systemic velocity as a function of the initial orbital separation. Orbits substantially expand under a Blaauw kick and receive systemic velocities of order tens of km s$^{-1}$. For systems with the tightest pre-SN orbits, systemic velocities may exceed 100 km s$^{-1}$, even without a SN kick.

Since the post-orbital parameters can be solved exactly, the merger time of the system in physical units can be determined as a function of mass parameters and the initial orbital separation only:
\begin{equation}
\tmerge = \frac{12}{19} \frac{1}{\beta_G} \frac{\beta^4}{(2\beta - 1)^4} H \left( e \right) a_i^4, \label{eq:blaauw_tmerge}
\end{equation}
where $e$ is determined from Equation \ref{eq:blaauw_e}. Combining this merger time with the systemic velocity provides an estimate of the distance traveled between the position of the secondary's core collapse and the merger location of the system ($d_{\rm travel} = \vsys \tmerge$) using the expressions for $\vsys$ and $\tmerge$ from Equations \ref{eq:blaauw_vsys} and \ref{eq:blaauw_tmerge}, respectively. Of course, this travel distance is calculated under the assumption that the system is isolated. 

The right two panels of Figure \ref{fig:blaauw} show the delay time and distance traveled as a function of initial orbital separation. Only those binaries with pre-SN orbital separations smaller than $\approx$3 \Rsun\ merge within a Hubble time. In an isolated environment, those binaries that merge may travel between a few kpc and a Mpc before doing so.

\section{Post-SN Orbital Characteristics}
\label{sec:orbital_characteristics}

When the SN kick is a non-negligible fraction of the orbital velocity, the distribution of kick directions and magnitudes must be taken into account. In this section, we discuss how the distribution of SN kicks affects the post-SN orbits.

\subsection{Survival Fraction}

\begin{figure}
	\includegraphics[width=\columnwidth]{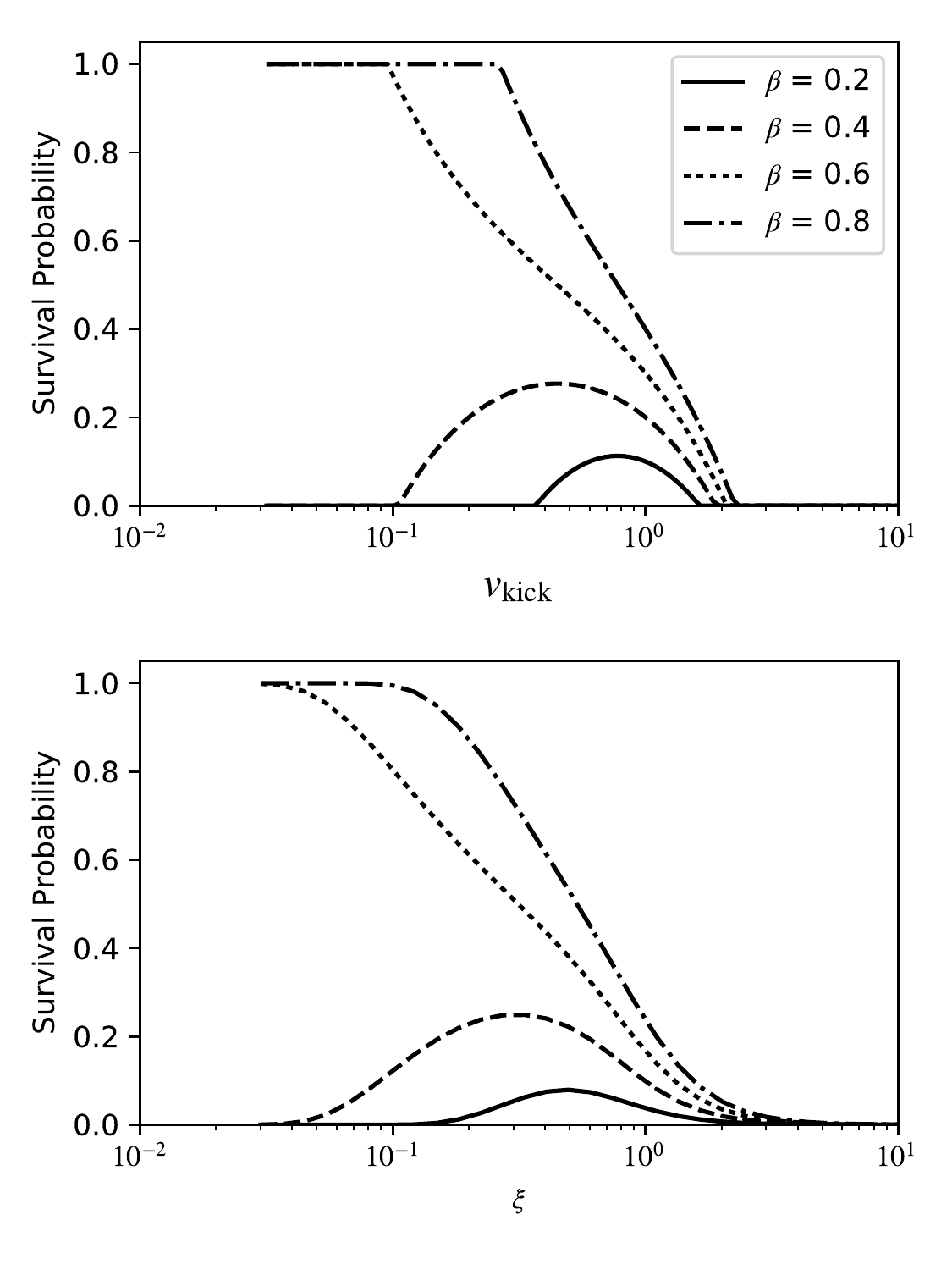}
    \caption{ The fraction of systems that remain bound after the SN, $P({\rm survive})$, as a function of the kick velocity $v_{\rm k}$ for single-magnitude kicks (top panel) or the non-dimensional dispersion velocity $\xi$ for a Maxwellian kick velocity distribution (bottom panel). Small $v_{\rm k}$ or $\xi$ represents the Blaauw limit in which only systems with $\beta$$>$0.5 survive. }
    \label{fig:survival_fraction}
\end{figure}

\begin{figure*}
	\includegraphics[width=\textwidth]{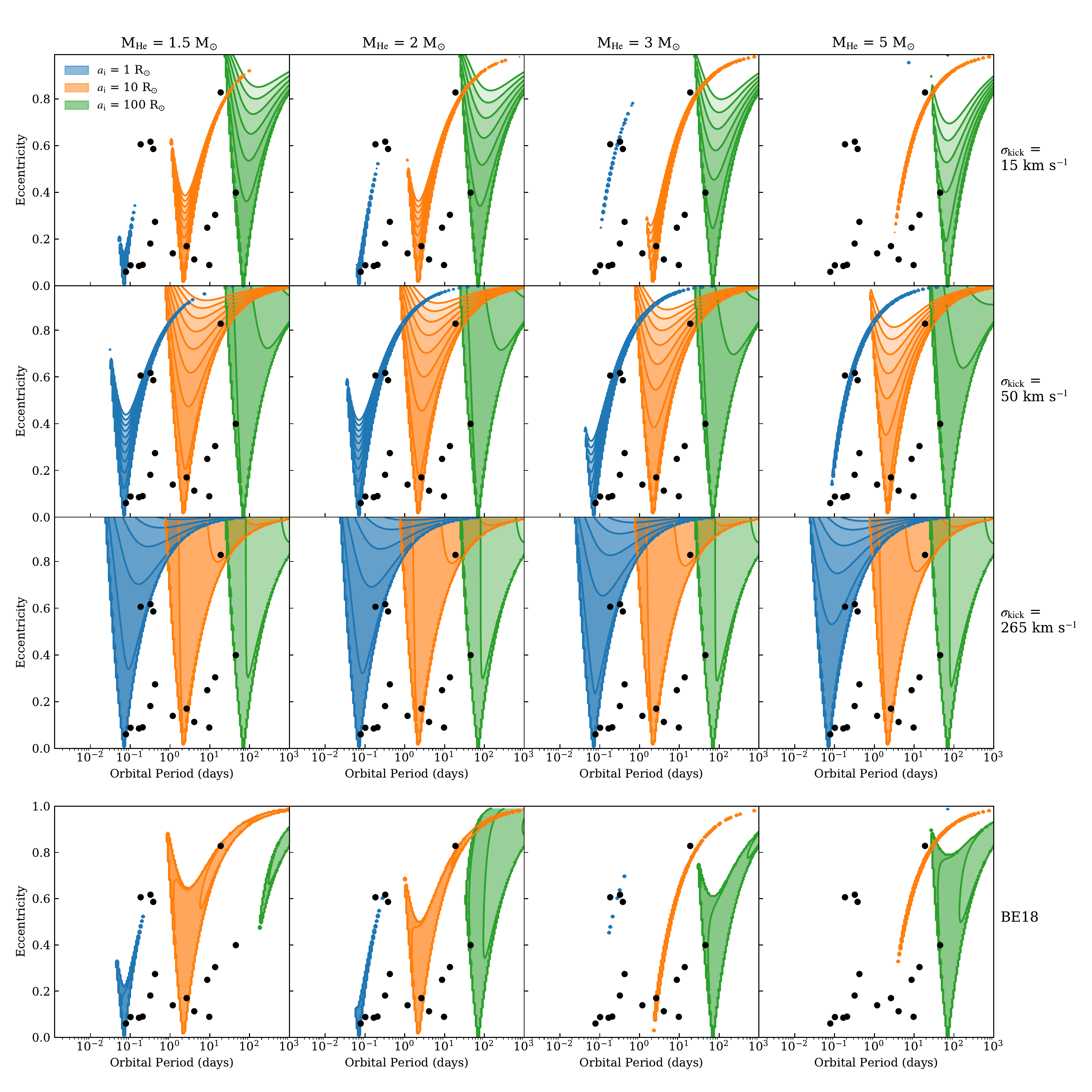}
    \caption{ The post-SN orbital period and eccentricity distribution for a grid in $\MHe$, $\sigma_k$, and $a_i$. Black points show the positions of the 17 known Galactic DNSs. Color contours are separated logarithmically. The DNSs in the Milky Way can be explained by pre-SN orbital separations ranging from 1 to 44 \Rsun\ irrespective of the NS kick velocity or $M_{\rm He}$ (see Section \ref{sec:implications_galactic_DNS}). For comparison with the non-dimensional distributions in Figure \ref{fig:alpha_ecc}, our choice of He-star masses of 1.5, 2, 3, and 5 \Msun\ correspond to $\beta$ values of 0.97, 0.82, 0.64, and 0.44. }
    \label{fig:P_orb_ecc}
\end{figure*}

Since $\alpha$ has no dependence on the azimuthal kick angle, $\phi$, for a single-magnitude kick, we calculate the survival fraction of a binary due to a SN kick by integrating over the domain for $\cos \theta$: 
\begin{equation}
P({\rm survival}) = \int_{-1}^{\cos \theta_{\rm max}} \dd \cos \theta\ P(\cos \theta),
\end{equation}
where from Equation \ref{eq:P_cos_theta}, $P(\cos \theta)=1/2$. We determine $\cos \theta_{\rm max}$ by solving Equation \ref{eq:alpha_def} for $\cos \theta$ and setting $\alpha\rightarrow \infty$:
\begin{equation}
\cos \theta_{\rm max} = \frac{2\beta - v_{\rm k}^2 - 1}{2 v_{\rm k}}. \label{eq:cos_theta_max}
\end{equation}
If the quantity on the r.h.s.\ of Equation \ref{eq:cos_theta_max} is less than -1, then $P({\rm survival}) = 0$. Likewise, if the quantity on the r.h.s.\ is greater than unity, then $P({\rm survival}) = 1$. The top panel of Figure \ref{fig:survival_fraction} shows the survival fraction of binaries as a function of kick velocity, for four separate choices of $\beta$. As the kick velocity becomes small, the system is reasonably approximated as a Blaauw kick and, as discussed in Section \ref{sec:blaauw}, systems with $\beta<0.5$ disrupt while systems with $\beta>0.5$ all survive. 

To obtain the overall survival fraction of systems with kicks drawn from a Maxwellian velocity, we take a different approach, integrating $P(\alpha, e)$ over $e$ and $\alpha$ for the domain containing bound binaries:
\begin{equation}
P({\rm survive}) = \int_0^1\ \dd e\ \int_{\frac{1}{1+e}}^{\frac{1}{1-e}}\ \dd \alpha\ P(\alpha, e).
\end{equation}
We provide the result of this integral, which we can evaluate numerically, in the bottom panel of Figure \ref{fig:survival_fraction} as a function of $\xi$ for the same test values of $\beta$. The distribution appears as a ``rounded'' version of the top panel for single velocity kicks; the tails of the Maxwellian distribution extend the survival probabilities to smaller and larger $\xi$. 

For both panels of Figure \ref{fig:survival_fraction}, nearly all systems disrupt when $V_{\rm k} >2 V_{\rm orb}$. We also note that for those systems with $\beta$$<$0.5, the maximum survival fraction is obtained when kick velocities are somewhat lower than the orbital velocities.

\subsection{Post-SN Orbital Period and Eccentricity}

\begin{figure*}
	\includegraphics[width=\textwidth]{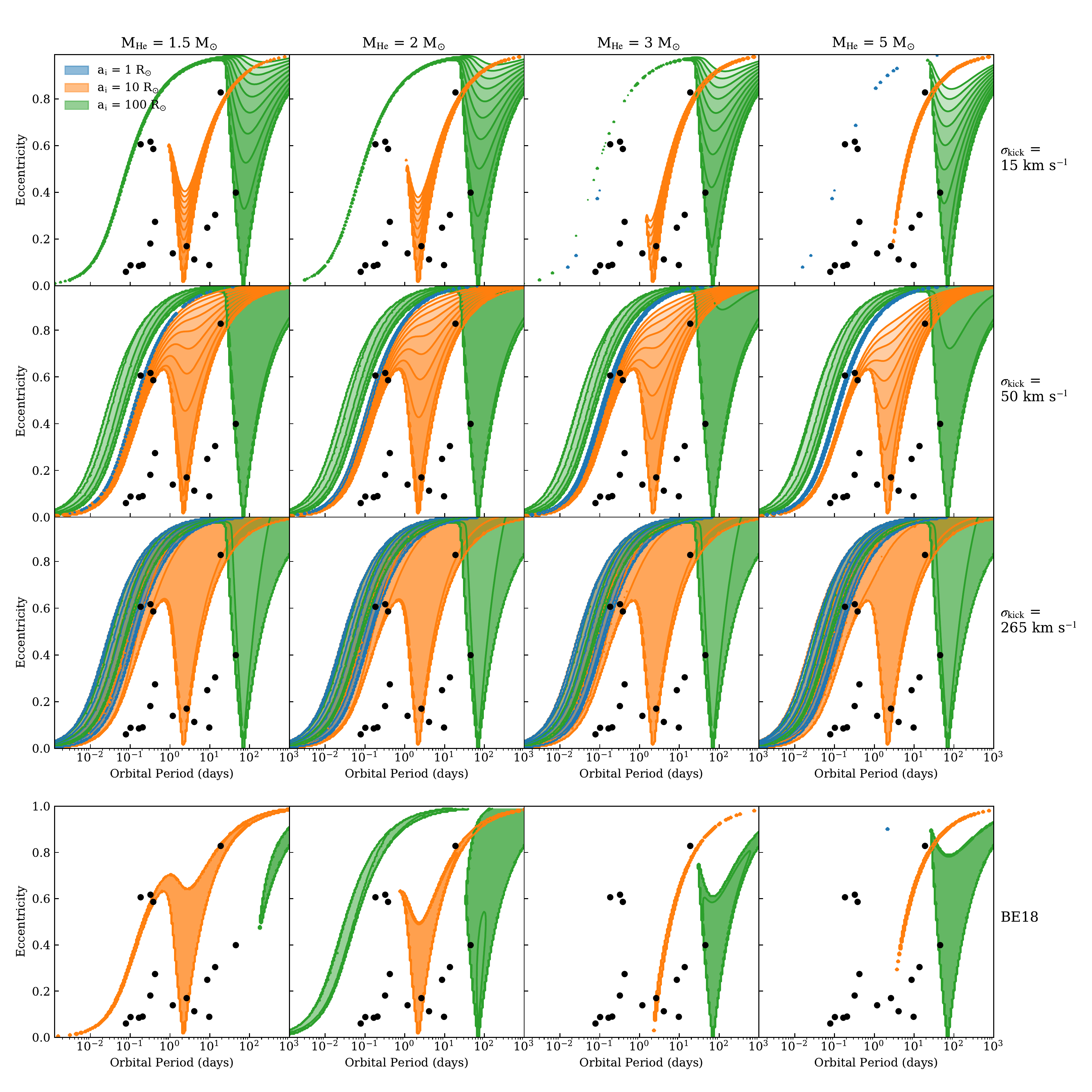}
    \caption{ The distribution of orbital periods and eccentricities of a population of DNSs after the orbits have evolved for 1 Gyr due to gravitational wave radiation. Systems that have merged are not shown. While orbital decay due to gravitational wave radiation produces a tail in the distribution at small orbital periods and eccentricities, this tail only starts to overlap the observed systems after a Hubble time. Even then, it contains very few systems. }
    \label{fig:P_orb_ecc_1Gyr}
\end{figure*}

Although Figure \ref{fig:alpha_ecc} shows the post-SN orbital separation and eccentricity for one particular value of $\beta$ and $\xi$, the range of realistic pre-SN orbits leads to a diverse set of possible post-SN orbital distributions. In Figure \ref{fig:P_orb_ecc}, we show the post-SN orbital distribution of a grid of models in physical units. The top three rows show the effects of Maxwellian kicks with $\sigma_k$ of 15, 50, and 265 km s$^{-1}$, while the columns show the models in which $M_{\rm He}$ = 1.5, 2, 3, and 5 \Msun. The bottom row shows the analogous distributions using the single-magnitude SN kick prescription from \citet{bray18} with velocities provided in Table \ref{tab:V_kick}. Blue, orange, and green colored contours (which have logarithmically separated contour lines) in each panel show the distributions for initial orbital separations of 1, 10, and 100 \Rsun. We use this grid scheme and color combination throughout this work. Note that since we plot \Porb\ in log-scale, Figure \ref{fig:P_orb_ecc} shows $P(\log \alpha, e)$ $=$ $\alpha$ $\times$ $P(\alpha, e)$. 

Higher density regions in the distribution are typically found near the limits of the distribution, near the $\alpha$$=$$(1 \pm e)^{-1}$ lines. Models with relatively higher $\MHe$ or larger kick velocities tend to shift the distribution to larger eccentricities, although the manner in which they do so is somewhat different; models with larger kicks stretch the entire distribution toward higher eccentricities, while models with larger $\MHe$ shift the distributions along the $\alpha$$=$$(1-e)^{-1}$ line. 

Finally, we overplot the positions of the 17 known field Galactic DNSs (black points). Regardless of the kick velocities or $M_{\rm He}$, Figure \ref{fig:P_orb_ecc} shows that the observed population is spanned by our choice of $a_{\rm i}$, suggesting that the DNSs in the Milky Way all had pre-SN orbital separations between $\sim$1 and $\sim$100 \Rsun. As we explicitly show in the following section, this conclusion is unaltered when the effects of gravitational wave radiation on these distributions are accounted for. We discuss the implications of this conclusion in Section \ref{sec:implications_galactic_DNS}.

\begin{figure*}
	\includegraphics[width=\textwidth]{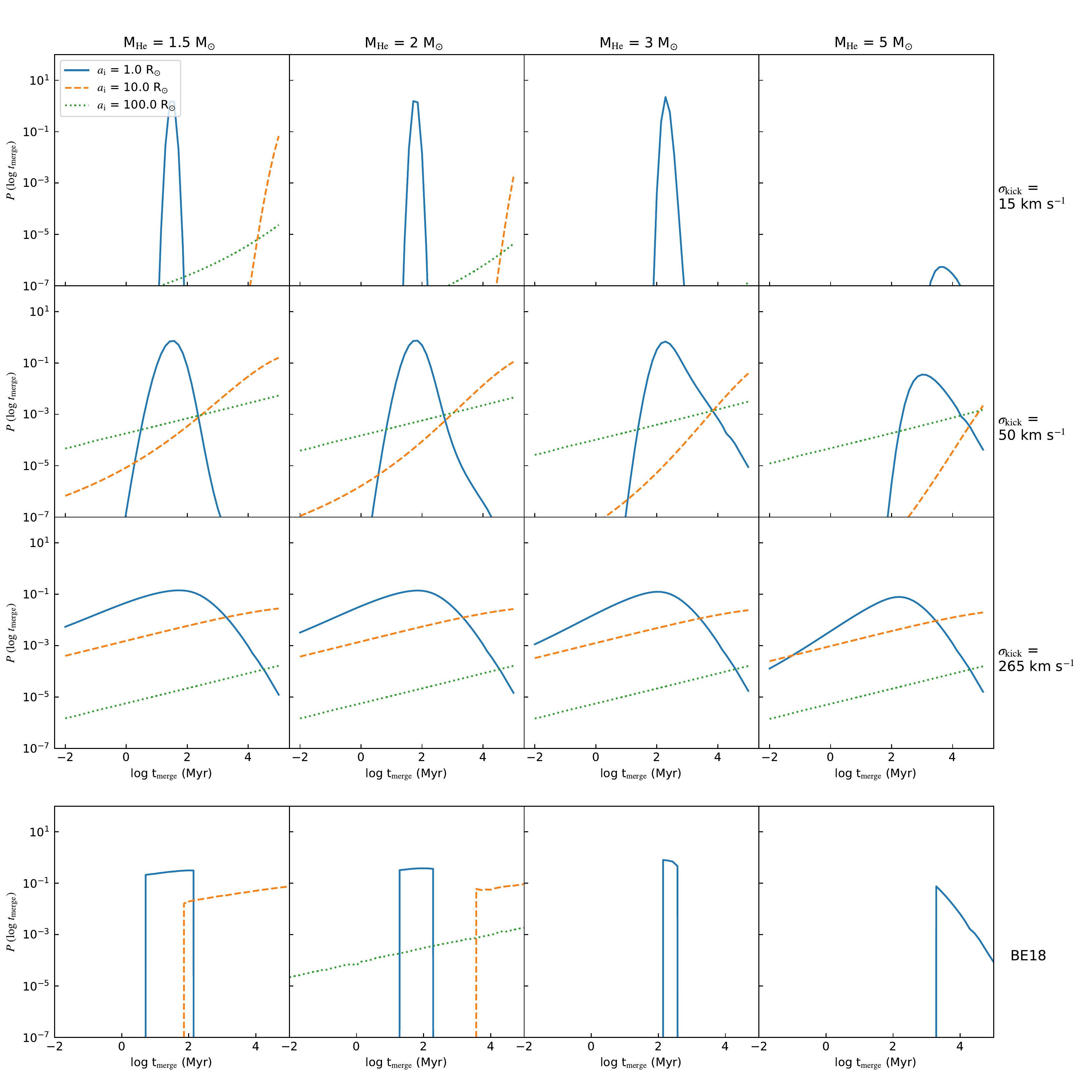}
    \caption{ The distribution of $\tmerge$ for a grid of models. Systems with $a_{\rm i}\sim 1$ \Rsun, merge within $\sim$100 Myr, while the steep dependence on orbital separation makes systems with $a_{\rm i}\sim 10$ \Rsun\ unlikely to merge within a Hubble time. Systems with low kick velocities and small $a_i$ have highly peaked distributions, which broadens substantially for more massive $\MHe$, wider pre-SN orbits, and larger $\sigma_k$. The green and orange lines are missing from certain panels as for these combinations of parameters either the merger times are beyond the plot limits or the systems nearly all disrupt during the second SN. Note that we define $\tmerge$ to be the time it takes the DNS to merge as measured from the second SN; an additional $\sim$tens of Myr should be added to obtain merger times measured with respect to the zero age main sequence (see Figure \ref{fig:DNS_formation_time}).}
    \label{fig:t_merge}
\end{figure*}

\subsection{Orbital Period and Eccentricity after Orbital Decay}

We now wish to evolve the distributions shown in Figure \ref{fig:P_orb_ecc} due to the general relativistic effects of orbital decay and circularization, as well as removing those systems that have merged. We can do this by converting Equation \ref{eq:P_alpha_ecc_GR} from non-dimensional to physical units and applying Kepler's third law to obtain orbital periods. Figure \ref{fig:P_orb_ecc_1Gyr} shows these distributions after the DNS orbits have evolved for one Gyr (i.e., one Gyr after a burst of star formation). Systems that have merged are not shown. 

Comparison with the unevolved orbital distributions in Figure \ref{fig:P_orb_ecc} shows that the effects from gravitational wave radiation on the orbits are profound. The distributions all show tails that extend toward smaller orbital periods and eccentricities, as systems circularize and decay. Nearly all of the systems with pre-SN orbits of 1 $\Rsun$ (blue distributions) have merged, with the exception of those systems that received large kicks; only a sliver in parameter space remains for these systems in the top two rows of panels, that requires a finer sampling grid than that used to generate Figure \ref{fig:P_orb_ecc} to be captured.

We elect not to provide analogous figures for populations of DNSs that have evolved for either less or more time, as for the purposes of this work the qualitative differences can be determined using Figures \ref{fig:P_orb_ecc} and \ref{fig:P_orb_ecc_1Gyr} as guides. Figure \ref{fig:alpha_ecc_GW} shows that as time passes, systems are progressively peeled away from the distributions starting from the top left. Using this trend as example, populations that have evolved for less than one Gyr look largely similar to those distributions in Figure \ref{fig:P_orb_ecc_1Gyr}, but with the characteristic, low-$\Porb$ tails found at even smaller orbital periods.

\section{Merger Times}
\label{sec:merger_time}

We next investigate the merger time distribution of the population of DNSs. Note that the merger time distributions described in this section refer to the time between the second SN and the DNS merger. 

The distribution of merger times can be determined by transforming $\alpha$ into $\taumerge$ in the bivariate $\alpha - e$ distribution, then integrating over all eccentricities:
\begin{equation}
P(\tau_{\rm merge}) = \int_0^1\ \dd e\ P(\alpha^*, e)\ \left| \frac{\partial \alpha}{\partial \tau} \right|_{\tau = \taumerge}, \label{eq:P_tau_merge}
\end{equation}
where $\alpha^*$ is the value of $\alpha$ for a given combination of $e$ and $\taumerge$ which can be determined by solving Equation \ref{eq:tau_merge} for $\alpha$. From this same equation, one finds the partial derivative term in Equation \ref{eq:P_tau_merge} is:
\begin{equation}
\left| \frac{\partial \alpha}{\partial \tau} \right|_{\tau = \taumerge} = \left[ \frac{19}{12288} \frac{1}{H(e)} \right]^{1/4} \taumerge^{-3/4}.
\end{equation}

\begin{figure}
	\includegraphics[width=\columnwidth]{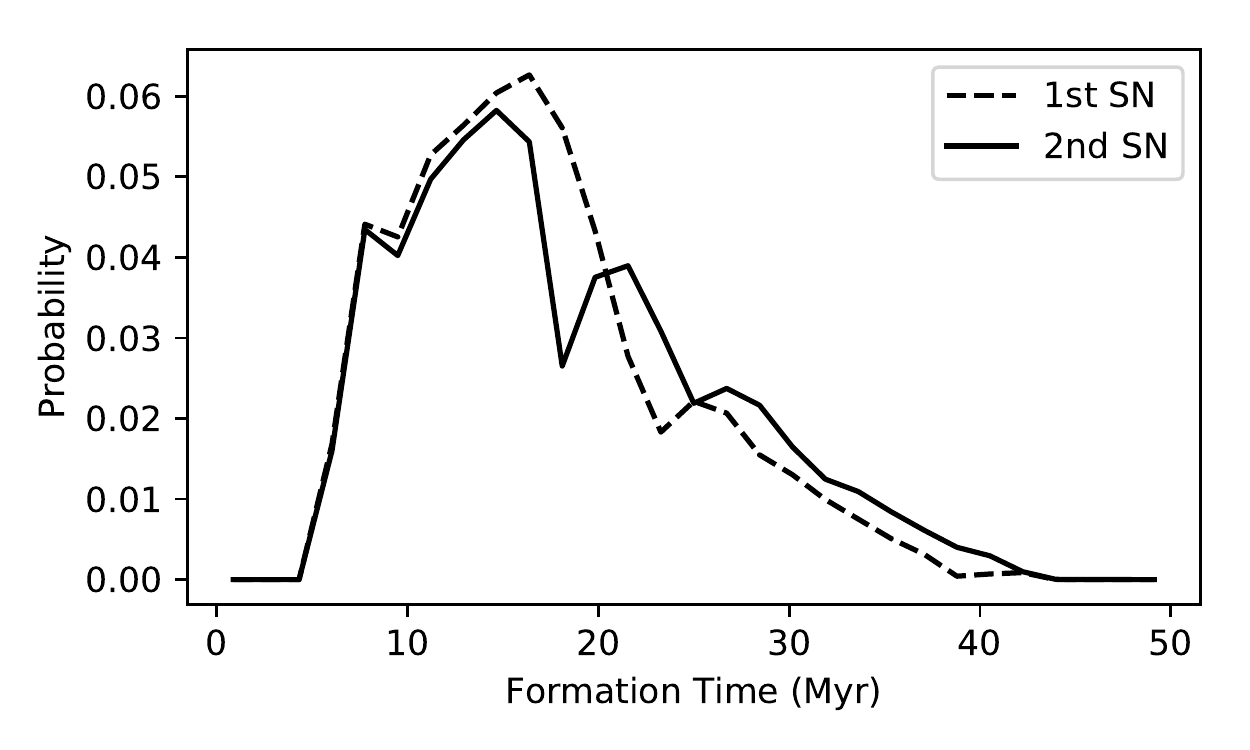}
    \caption{ Theoretical predictions for the formation time of DNSs. The first SN typically occurs 10-20 Myr after a star formation episode, and the second SN shortly ($\lesssim$3 Myr) afterwards. }
    \label{fig:DNS_formation_time}
\end{figure}

Figure \ref{fig:t_merge} shows the distribution of merger times for our grid of models in $\sigma_{\rm k}$, $M_{\rm He}$, and $a_{\rm i}$. Most systems with $a_{\rm i}\sim 1$ \Rsun\ merge within 10-100 Myr after the second SN, regardless of other parameters. The steep dependence of the merger time on $a_{\rm i}$ means that even systems with pre-SN orbital separations $\sim$10 \Rsun\ are unlikely to merge within a Hubble time. Note that this result applies regardless of whether the kick has a single magnitude, as in the bottom row \citep[representing the kick prescription from][]{bray18}, or the top three rows (representing a Maxwellian kick distribution) of Figure \ref{fig:t_merge}. Furthermore, systems with the lowest kick velocities (with respect to $V_{\rm orb}$) produce highly peaked merger time distributions; these are well approximated by a Blaauw kick, and the NS natal kick can be safely ignored. Increasing the kick velocity on the other hand, broadens the merger time distribution. 

Using the results from Section \ref{sec:blaauw}, we can determine the position of the sharp peak in merger times in the limit of low $\xi$. Therefore, in the limit of low kick velocities (compared to the pre-SN orbital velocity) and for $\beta$$\gtrsim$0.57 (from the limit on Equation \ref{eq:H_e_low} that $e$$<$0.75), Equation \ref{eq:blaauw_tmerge} can be approximated in non-dimensional units as:
\begin{equation}
\tau_{\rm merge} \approx \frac{1}{\sqrt{2\beta -1}}\frac{1}{\beta^3}. \label{eq:tau_merge_approx}
\end{equation}
To calculate the merger time for 0.5$<$$\beta$$<$0.57, the full expression for $H(e)$ in Equation \ref{eq:H_e} must be used in Equation \ref{eq:blaauw_tmerge} (again for $\beta$$<$0.5, all systems in the low-kick limit disrupt).

We provide the merger time in physical units adopting $M_2$$=$$\MNS$$=$1.4\Msun:
\begin{eqnarray}
t_{\rm merge} &\approx& 30 \left( \frac{a_i}{1 \Rsun} \right)^4 {\rm Myr} : \MHe = 1.5 \Msun \\
t_{\rm merge} &\approx& 60 \left( \frac{a_i}{1 \Rsun} \right)^4 {\rm Myr} : \MHe = 2 \Msun \\
t_{\rm merge} &\approx& 200 \left( \frac{a_i}{1 \Rsun} \right)^4 {\rm Myr} : \MHe = 3 \Msun. 
\end{eqnarray}
We want to reiterate that the above approximations are only valid when the kick velocities are a small fraction of the orbital velocity. 

The merger times above refer to how quickly after the second SN the system will merge. We use our recently developed binary population modeling code {\tt dart\_board} \citep{andrews18} to form an estimate for the distribution of times prior to the first and second SNe forming DNSs. Rather than Monte Carlo random sampling as is used in BPS, {\tt dart\_board} uses a Markov-chain Monte Carlo approach to model the formation of stellar binaries. We define the likelihood function so that {\tt dart\_board} only focuses on DNSs, and we use a flat prior on the birth time (constant star formation history). We use a version of {\tt BSE} \citep{hurley02} to rapidly evolve stellar binaries, with modifications described by \citet{kiel08} and \citet{rodriguez16}. 

Figure \ref{fig:DNS_formation_time} shows that the first SN typically occurs 10-20 Myr after a star formation episode, with the second SN occurring shortly afterwards. We use standard binary evolution parameters in this model. While differences in the assumptions about binary evolution physics, particularly that related to mass transfer, may lead to large differences in the distribution of DNSs, it is unlikely that the DNS formation time distribution in Figure \ref{fig:DNS_formation_time} will significantly vary. To produce the distribution of merger times after a star-formation episode, one should combine the distribution of times to the second SN with the distribution of merger times in Figure \ref{fig:t_merge}. 

\begin{figure*}
	\includegraphics[width=\textwidth]{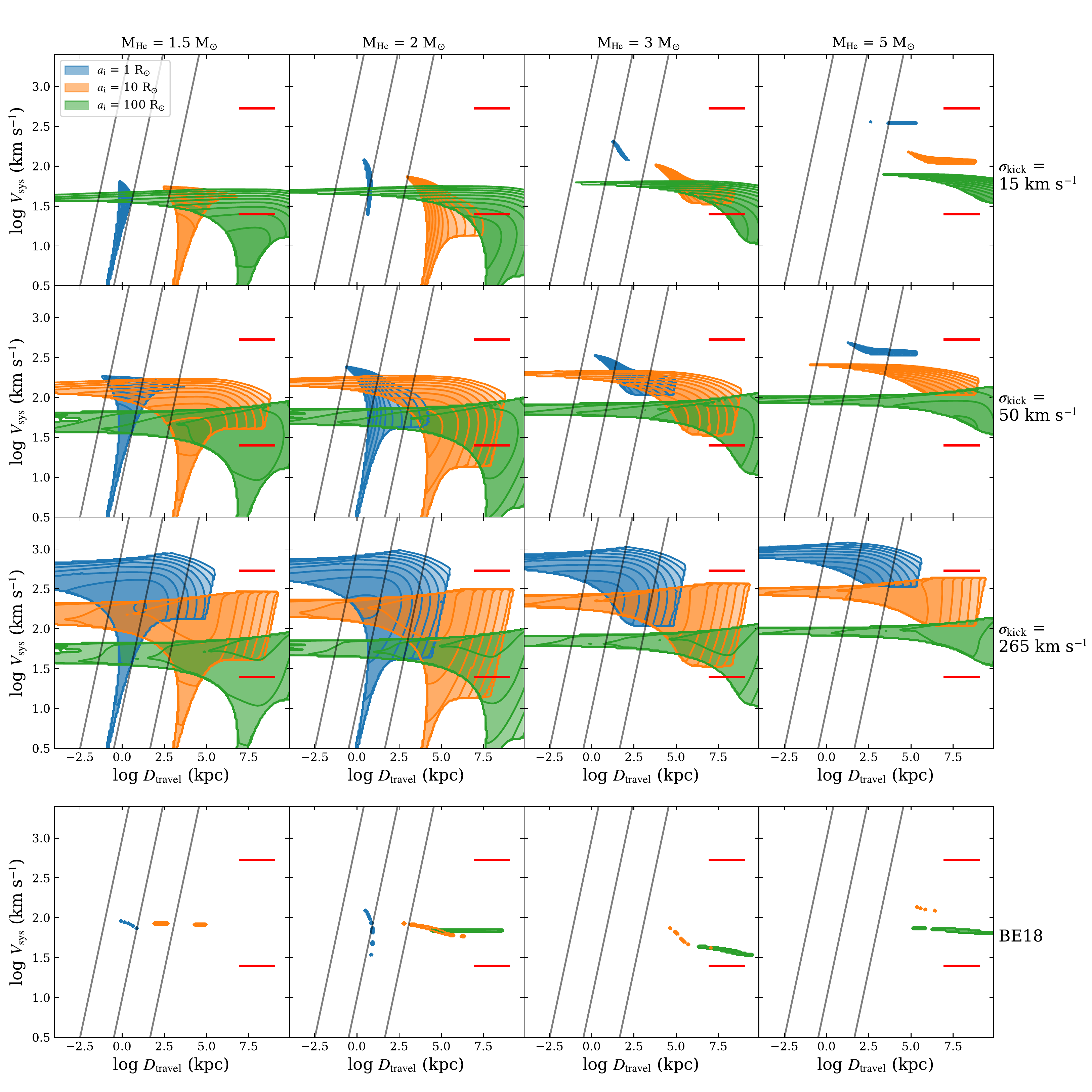}
    \caption{ The bivariate $P(V_{\rm sys}, D_{\rm travel})$ distribution for a grid of $\xi$ and $\beta$ values. Systems with initially smaller pre-SN orbits tend to have larger $V_{\rm sys}$ and smaller $D_{\rm travel}$. Black lines indicate constant merger times of 1 Myr, 100 Myr, and 14 Gyr, from left to right; any system to the right of the right-most black line takes longer than a Hubble time to merge. Depending on the combination of $\MHe$, $\sigma_k$, and $a_i$ produced by binary evolution for DNSs, systems may travel as little as $\lesssim$kpc or as much as $\sim$Mpc before merging. The distributions in the bottom panel (adopting the SN kick prescription from \citet{bray18}) are extremely narrow; the gaps that appear are numerical, resulting from our choice of grid spacing. Horizontal red lines in each panel indicate the local Milky Way escape velocity \citep[533 km s$^{-1}$][]{piffl14} and the typical escape velocity of an ultra-faint dwarf galaxy (25 km s$^{-1}$).} 
    \label{fig:V_sys_D_travel}
\end{figure*}

\section{Travel Distances and Systemic Velocities}
\label{sec:distance_velocity}

To determine the distance a system travels (ignoring the effects of an external gravitational potential), we can calculate the bivariate distribution $P(v_{\rm sys}, \dtravel)$, where $\dtravel$ is a non-dimensional travel distance:
\begin{equation}
\dtravel = \frac{4 \beta_G}{V_{\rm orb} a_i^4} D_{\rm travel} = \tau_{\rm merge} v_{\rm sys}. \label{eq:d_travel}
\end{equation}
We calculate $P(v_{\rm sys}, \dtravel)$ by first evaluating $P(\alpha, e, v_{\rm sys})$ from Equation \ref{eq:P_alpha_e_vsys}, multiplying by a Jacobian term to transform from $\alpha$ to $\dtravel$, then integrating over all $e$:
\begin{equation}
P(v_{\rm sys}, \dtravel) = \int_0^1 \dd e\ P(\alpha^*, e, \vsys) \left| \frac{\partial \alpha}{\partial d} \right|_{d = \dtravel} \label{eq:P_v_sys_d_travel}
\end{equation}
where $\alpha^*$ is determined by
\begin{equation}
\alpha^* = \left[ \frac{19}{48} \frac{1}{H(e)} \frac{\dtravel}{\vsys} \right]^{1/4}. \label{eq:alpha_star}
\end{equation}
Taking the partial derivative of this equation also provides the Jacobian term in Equation \ref{eq:P_v_sys_d_travel}:
\begin{equation}
\left| \frac{\partial \alpha}{\partial d} \right|_{d = \dtravel} = \frac{1}{4} \frac{\alpha^*}{\dtravel}. \label{eq:d_alpha_d_d_travel}
\end{equation}

We translate the non-dimensional distributions to physical units in Figure \ref{fig:V_sys_D_travel}. Our grid of models and color scheme is identical to that of Figure \ref{fig:P_orb_ecc}. Focusing on any individual panel in Figure \ref{fig:V_sys_D_travel}, we see that all other parameters being identical, those systems with relatively smaller pre-SN orbital separations tend to produce systems with larger $V_{\rm sys}$, shorter $\tmerge$, and shorter travel distances, $D_{\rm travel}$. For those systems with the tightest pre-SN orbits, $V_{\rm sys}$ is of order 10$^2$ km s$^{-1}$, but may reach somewhat higher velocities for sufficiently large kicks. 

Systems with larger $\MHe$, and therefore more mass lost during the SN, have larger $D_{\rm travel}$ and higher $V_{\rm sys}$, as can be seen when moving from left to right along the rows of panels in Figure \ref{fig:V_sys_D_travel}. Likewise, moving down along the columns of panels shows that increasing the kick velocities shifts the distributions toward larger systemic velocities, while only slightly shifting the distribution of distances that systems travel before merging.

Black lines indicate constant merger times of 1 Myr, 100 Myr, and 14 Gyr; systems to the right of the right-most line in each panel do not merge in a Hubble time. We extend the plot range to large $D_{\rm travel}$ to explicitly show that the distributions for DNSs with $a_{\rm i}$ of 10 \Rsun\ and 100 \Rsun\ peak beyond the observable range. 

Figure \ref{fig:V_sys_D_travel} shows that the distances traveled by a system are a strong function of the pre-SN orbital separation. For DNSs with $a_{\rm i}=1$ \Rsun, DNSs will typically travel 1-10 kpc before merging. Since systems with larger kick velocities have a broader range of merger times, they also have a correspondingly broader range of distances traveled before merging. For systems with either large kick velocities ($\sigma_k$$\approx$265 km s$^{-1}$) or large He-star masses (M$_{\rm He} > 3$ \Msun), these DNSs can leave galaxies with escape velocities of order several hundreds of km s$^{-1}$ (such as the Milky Way).

On the other hand, DNSs with $a_{\rm i}=10$ \Rsun\ can travel out to Mpc distances before merging (but almost certainly no larger than $\sim$10 Mpc), provided they indeed merge within a Hubble time. However, systems with $a_{\rm i}=10$ \Rsun\ typically do not have systemic velocities large enough to escape the gravitational potential of their host stellar population. Only for DNSs born with large kick velocities can these DNSs receive systemic velocities larger than $\sim$50 km s$^{-1}$. Note that systems with larger pre-SN separations can also travel large distances, but these typically will not merge within a Hubble time.

\section{Comparison with Observations}
\label{sec:discussion}

In this work we focus only on the effects of the supernova kick applied to NSs at birth and the subsequent orbital decay due to gravitational wave radiation. This approach allows us to generate expectations for key observables in both the population of Galactic DNSs as well as the population of DNS mergers detectable by the global network of gravitational wave observatories. Importantly, by ignoring the complex and uncertain binary evolution physics involved in the formation of DNSs prior to the SN forming the second NS, we can identify the range of possibilities of observable DNSs. However, this comes at a cost: ignoring the prior binary evolution restricts us from modeling the complete DNS population. For instance, we only test models with $a_{\rm i}$ of 1, 10, or 100 \Rsun\ when Nature produces a superposition of a continuous distribution of $a_{\rm i}$ spanning this range. Nevertheless, our results have applicability to a broad range of astrophysical subjects, which we cursorily outline below.

\subsection{Galactic DNSs}
\label{sec:implications_galactic_DNS}

\citet{flannery75} first showed that $\alpha$ is restricted to range between $(1+e)^{-1}$ and $(1-e)^{-1}$, which immediately places a constraint on the pre-SN orbital separation when given the post-SN orbital separation. Figure \ref{fig:P_orb_ecc} shows that this constraint can be quite stringent, especially for low $e$ systems. Even though we only observe the orbital separations of Galactic DNSs after their orbits have evolved for some time by gravitational wave radiation, the strong scaling of the merger time with orbital separation means that, as stated by \citet{peters64}, ``the system spends most of the time in a decay state for which a$_{\rm o}$$\approx$$a_{\rm f}$,'' where we have substituted our notation for his. Therefore, the system's orbital separation and eccentricity have most likely evolved little since the second SN.

Furthermore, even if DNS orbits have evolved substantially since their birth, we can still place a constraint on the pre-SN orbital separation, by analyzing how GR tends to shrink and circularize orbits. Blue dashed lines in the Figure \ref{fig:alpha_ecc} show the path traveled by an orbit in $\alpha-e$ parameter space over time, which \citet{peters64} express mathematically as (using the non-dimensional $\alpha$ instead of the orbital separation):
\begin{equation}
\left< \frac{\dd e}{\dd \alpha} \right>_{\rm GW} = 
\frac{19}{12}
\frac{e (1-e^2)}{\alpha}
\frac{1 + (121/304) e^2}{1 + (73/24) e^2 + (37/96) e^4}
\label{eq:de_dalpha_GW}
\end{equation}
The corresponding slope along the boundary line $e = 1 - 1/\alpha$ is:
\begin{equation}
\left( \frac{\dd e}{\dd \alpha} \right)_{\rm SN} = \alpha^2.
\end{equation}
By inserting $1 - 1/\alpha$ for $e$ in Equation \ref{eq:de_dalpha_GW} above, one finds that along the border $\alpha = (1-e)^{-1}$,
\begin{equation}
\left( \frac{\dd e}{\dd \alpha} \right)_{\rm SN} > \left< \frac{\dd e}{\dd \alpha} \right>_{\rm GW}
\end{equation}
over the entire domain of $e$, at least to $\mathcal{O}(e^2)$. 

This result has an important implication: orbital decay due to gravitational wave radiation cannot shift a post-SN binary system from the allowable region in ($\alpha$,$e$) parameter space (bound by $e=1 - 1/\alpha$ on the higher $\alpha$ side) to a position to the right of this line (i.e. $e \nless 1 - 1/\alpha$, for $\alpha$$>$1). Therefore for any DNS formed through isolated binary evolution from an instantaneous kick in a circular orbit, we find a strict constraint on the {\it pre-SN} orbital separation: $a_i$$>$$(1-e)$$a$. The widest known DNS, J1930$-$1852 \citep{swiggum15}, has a current orbital separation of $\approx$73\Rsun\ and eccentricity of 0.399, which immediately implies a pre-SN orbital separation no less than 44 \Rsun. In some cases the inclination angles of DNSs are unknown, and only the projected orbital separation is known. Since $a_{\rm proj}$$\leq$$a$, the pre-SN orbital separation constraint holds even when using the projected separation: $a_i$$\geq$$(1-e)$$a_{\rm proj}$.   

At the same time, the shortest orbital period DNS, J1946$+$2052 \citep{stovall18}, has an orbital separation of 1 \Rsun\ and an eccentricity of 0.06. This implies that the pre-SN orbital separations of DNS progenitors range from as small as 1 \Rsun\ to at least as large as 44 \Rsun. Orbital decay due to gravitational wave radiation does not affect this conclusion since, although Figures~\ref{fig:alpha_ecc_GW} and \ref{fig:P_orb_ecc_1Gyr} show that orbital decay due to GR produces a small orbital separation tail at low eccentricities, the chance of observing a system as it passes through this region is extremely unlikely. DNSs may form with pre-SN orbital separations even shorter than 1 \Rsun, but these would merge very quickly (within tens of Myr) after formation.

\subsection{Short GRBs}

The accompaniment of a short GRB with the gravitational wave merger event GW170817 \citep{ligo_detection} allows us to attribute the formation of at least some short GRBs to the merger of two NSs in a DNS \citep{fong17}. Since we have detected only one DNS merger and many hundreds of GRBs \citep[for a review, see][]{berger11}, currently the population of short GRBs may provide the best observational insight into the properties of DNS mergers. For instance, by analyzing the star formation history of GRB host galaxies, \citet{leibler10} derive a delay time distribution of short GRBs to be $\sim$200 Myr for late-type galaxies and $\sim$3 Gyr for early-type galaxies \citep[see also][]{wanderman15, zhang18}. Figure \ref{fig:t_merge} shows that this range can be easily explained by DNS mergers using a reasonable range of SN kick distributions, pre-SN masses, and pre-SN orbital separations. 

As a further example of how short GRBs can constrain DNS formation, \citet{fong13} find that, of a sample of 22 GRBs, $\approx$20\% are found more than 5 galactic radii away from their host galaxies. Other GRB samples have similarly included a fraction of events that are found at large distances from their host galaxy \citep{prochaska06, fong10} or even so far away that the host galaxy cannot be definitively identified \citep{fong13b}. The authors of these observational studies have tended to attribute the distances traveled to SN kicks, an idea further explored in multiple theoretical works \citep{bagot98, bloom99, bulik99, perna02, niino08, zemp09, kelley10, behroozi14,wiggins18}. Figure \ref{fig:V_sys_D_travel} indicates that only systems with either large kicks ($\sim$10$^2$ km s$^{-1}$) or large pre-SN masses ($M_{\rm He}\gtrsim 5$ \Msun) can escape Milky Way-mass galaxies. These figures suggest that better statistics on the offset distribution of short GRBs from their galaxy hosts can constrain DNS formation scenarios.

\subsection{GW170817 and Future Detections of DNS Mergers}
\label{sec:implications_LIGO}

In modeling the DNS forming GW170817, \citet{ligo_progenitor} find typical delay times between formation and merger of $\sim$4 Gyr (although the dispersion around this age is large). Regardless of whether NGC 4993, the host galaxy to GW170817, had some recent star formation \citep{ebrova18} or not \citep{blanchard17, Im17, levan17}, our analysis in Section \ref{sec:merger_time} finds that the formation, evolution, and merger of the DNS forming GW170817 can easily accommodate either a stellar population age of $\sim$10$^2$ Myr or $>$10$^3$ Myr. Based on Figure \ref{fig:t_merge}, for a system to merge with a delay time longer than $\sim$1 Gyr, we find the system likely had a {\it pre-SN} orbital separation between 1 and 10 \Rsun. This is in agreement with the results of \citet{blanchard17} who derive a {\it post-SN} orbital separation of $\approx$4.5 \Rsun\ after adopting a merger time of $\approx$11.2 Gyr, derived from their star formation history of NGC 4993. 

The presence of an electromagnetic (EM) counterpart to GW170817 that was 2 kpc away from its host galaxy, NGC 4993 \citep{ligo_progenitor}, gives us the tantalizing possibility that, with an increased sample size in the future, we may be able to observationally constrain the distribution of distances traveled by DNSs before merging, similar to the offset distributions of short GRBs. For pre-SN orbital separations of $\sim$1\Rsun, travel distances are of order kpc, in agreement with the projected separation of 2 kpc between the EM counterpart to GW170817 and NGC 4993 (this is a crude comparison since (a) the DNS forming GW170817 likely orbited around NGC 4993 many times before merging, and (b) it could have formed in the outskirts of NGC 4993, rather than the bulge). However, for systems with pre-SN orbital separations $\sim$10 \Rsun, if the systems escape their galaxy hosts, they typically travel $\sim$Mpc before merging; the only limitation is the finite age of the Universe. Of course, the situation is complicated by the gravitational potential of a particular system's host galaxy, which may evolve over cosmological timescales \citep[e.g.,][]{zemp09}. 

We therefore expect that some fraction of DNS mergers detected by gravitational wave observatories may have EM counterparts far from any galaxy (similar to some short GRBs), or even near the wrong galaxy (at least in projection). Strategies for multi-wavelength follow-up to identify the EM counterpart for any particular gravitational wave event ought to consider this possibility. Furthermore, this could complicate constraints on the expansion of the Universe using DNS mergers as these constraints require the definitive association with a particular host galaxy.

\subsection{Ultra-faint Dwarf Galaxies}
\label{sec:implications_r_process}

The detection of $r$-process enhanced, metal-poor stars \citep{sneden94, hill02} implies that whatever the astrophysical site of the $r$-process, it must be active early in the age of the Universe and must be a ``local'' nucleosynthesis event. Observations of GW170817 in both gravitational waves and optical spectroscopy have identified DNS mergers as a major $r$-process site \citep{cowperthwaite17,nicholl17,chornock17}. Furthermore, the presence of $r$-process enriched stars in ultrafaint dwarf galaxies (UFDs) suggests that even galaxies of low mass can retain $r$-process enriched material once it is created \citep{ji16a, ji16b,hansen17, ji18}.

Assuming DNSs to be the site of $r$-process enrichment in UFDs, the presence of chemically peculiar stars in these galaxies has two important implications for the merger of DNSs. First, DNSs must have formed and merged early in the lifetime of the Universe \citep{ji16a}, before the second round of stars formed $\sim$100 Myr after the first stars were formed \citep{ji15}. Second, DNSs must merge within their host galaxies for the resulting $r$-process material to be retained \citep{safarzadeh17}. Using BPS, \citet{safarzadeh18} argue that DNSs mergers may be responsible for the observed $r$-process enrichment of two specific UFDs, Reticulum II and Tucanae III, if DNSs undergo a phase of unstable Case BB mass transfer. In this case, despite their large escape velocities, the DNSs merge quickly after formation before they have traveled very far.

Using our methodology, we can more rigorously determine what is required to form and merge DNSs quickly enough to enrich the next generation of stars in UFDs. Figure \ref{fig:t_merge} shows that only systems with pre-SN orbital separations of $\sim$1 \Rsun\ can merge within $\sim$10$^2$ Myr. Figure \ref{fig:V_sys_D_travel} shows that even those systems that have large enough systemic velocities to escape an UFD will likely merge within $\sim$10 kpc, for He-star masses $\lesssim$3 \Msun\ regardless of the kick velocity. For smaller $M_{\rm He}$, this travel distance decreases even more. Although a complete quantitative analysis including merger rates is outside the scope of this work, our results suggest that DNSs are qualitatively capable of causing the $r$-process enrichment of UFDs.

\section{Conclusions and Future Directions}
\label{sec:conclusions}

We disentangle the effects of binary evolution physics from the SN kick forming the second NS in a DNS system, followed by its inspiral due to gravitational wave-induced orbital decay. This allows us to gain an intuition for how uncertain physics involved in binary evolution may affect the conclusions for DNS physics, in particular the travel distances, merger times, and systemic velocities of merging DNSs. 

Assuming instantaneous, isotropic SN kicks from either a single-magnitude or a Maxwellian distribution of kicks within a circular pre-SN binary, we show how the post-SN distribution of eccentricities and orbital periods is populated based on different assumptions about the pre-SN orbital configurations. We compare to the observed population of DNSs, determining that they were formed from pre-SN orbital separations ranging from 1 to 44 \Rsun. 

We calculate the delay time distribution of DNSs, measured with respect to the formation of the second SN. The range of pre-SN orbital separations indicated by the Galactic DNSs produces delay times ranging from tens of Myr to larger than a Hubble time. For small kick models, typically only those systems with the tightest pre-SN orbits merge, whereas with large kick models, a broad range of merger times is possible. 

We provide distributions describing the systemic velocity received by a newly born DNS and the distance traveled before the system merges. Systems typically travel no faster than the pre-SN orbital velocity, and often significantly slower. In the low kick limit, only the tightest pre-SN orbits will receive a large enough kick velocity to escape their host galaxies, and then only in small gravitational potentials. For larger kick velocities, systems may travel much larger distances -- as large as $\sim$Mpc -- before merging. As discussed by previous authors, this provides an explanation for the so-called ``host-less'' GRBs that are observed far from any galaxy.

The semi-analytic methods here may be combined in the future with binary population synthesis simulations that can produce realistic distributions of pre-SN orbital configurations. As the number of detections increases, compact object merger events can start to be used to constrain theoretical binary evolution models \citep[for example, see work by][]{mandel17, barrett18}. In particular, a significant sample of DNS travel distances (as measured by the distance between the EM counterpart to future LIGO/Virgo merger events and the DNS host galaxies) would greatly help constrain the distribution of pre-SN orbital configurations using the results presented here.

\section*{Acknowledgements}

We thank the anonymous referee whose detailed comments greatly improved the quality of the manuscript. The authors would also like to acknowledge useful discussions with Vicky Kalogera, Mike Zevin, Katie Breivik, Enrico Ramirez-Ruiz, and Ilya Mandel. The authors acknowledge funding from the European Research Council under the European Union's Seventh Framework Programme (FP/2007-2013)/ERC Grant Agreement No 617001. This project received funding from the European Union's Horizon 2020 research and innovation programme under the Marie Sklodowska-Curie RISE action, grand agreement No 691164 (ASTROSTAT).

\bibliographystyle{mnras}
\bibliography{references}

\bsp	
\label{lastpage}
\end{document}